\documentclass[prl,aps,twocolumn,showpacs,10pt]{revtex4-1}
\usepackage{color}
\usepackage{amsmath}
\usepackage{amssymb}
\usepackage{graphicx}
\usepackage{bm}
\usepackage[normalem]{ulem}
\graphicspath{{figures/}}

\begin{document}

\title{Antiphase Synchronization in a Flagellar-Dominance Mutant of {\it Chlamydomonas}}
\author{Kyriacos C. Leptos$^{1}$, Kirsty Y. Wan$^{1}$,  Marco Polin$^{1}$, 
Idan Tuval$^{2}$, Adriana I. Pesci$^{1}$ and 
Raymond E. Goldstein$^{1}$}
\affiliation{$^{1}$Department of Applied Mathematics and Theoretical
Physics, 
%Centre for Mathematical Sciences, 
University of Cambridge, Wilberforce Road, Cambridge CB3 0WA, 
United Kingdom\\ $^{2}$Mediterranean Institute for Advanced Studies (CSI-UIB), E-07190 Esporles, Spain}
\date{\today}

\begin{abstract}
Groups of beating flagella or cilia often synchronize so that neighboring 
filaments have identical frequencies and phases.  A prime example is provided by the 
unicellular biflagellate {\it Chlamydomonas reinhardtii}, which typically displays synchronous \textit{in-phase} beating 
in a low-Reynolds number version of breaststroke 
swimming. We report here the discovery that \textit{ptx1}, a flagellar dominance mutant of \textit{C. reinhardtii},
can exhibit synchronization in precise \textit{antiphase}, as in the freestyle swimming stroke.  
Long-duration high-speed imaging shows that \textit{ptx1} flagella switch stochastically between in-phase and
antiphase states, and that the latter has a distinct waveform 
and significantly higher frequency, both of which are strikingly similar to those found during phase slips that
stochastically interrupt in-phase beating of the wild type. Possible mechanisms underlying these observations are discussed.
\end{abstract}


\maketitle

Living creatures capable of motion seldom restrict themselves to a single mode of self-propulsion. 
Pairs of appendages of multilegged
organisms can be actuated synchronously in-phase, synchronously out-of-phase, or asynchronously, 
typically under neuronal control through a ``central pattern generator\rq\rq{}\cite{Collins1993}. 
In the world of aquatic microorganisms, where there is no direct analog of a central nervous system, the 
cilia and flagella adorning algae and bacteria are the ``limbs'' which exhibit various sychronization modes, 
generating swimming \cite{LaugaGoldstein2012}. 
Within a given eukaryotic organism, the undulations of flagella which arise from molecular motors distributed
along the length of the filaments can be found to synchronize in two stereotypical ways.  
Biflagellate cells epitomized by the alga {\it Chlamydomonas} \cite{chlamy_sourcebook}
display synchronous beating with identical frequencies and phases \cite{RNall,Polin}.   Those with
multitudes of cilia or flagella, such as unicellular {\it Paramecium} \cite{KnightJones}
or multicellular {\it Volvox} \cite{Brumley}, exhibit metachronal waves in which flagella with a common
frequency have phases that vary monotonically with position.
Theory \cite{RaminGeneric,Niedermayer,Guirao} suggests that these modes of synchronization can arise from  
fluid dynamical coupling between flagella, possibly assisted by waveform compliance.
 
\begin{figure}[b]
\includegraphics[width=0.9\columnwidth]{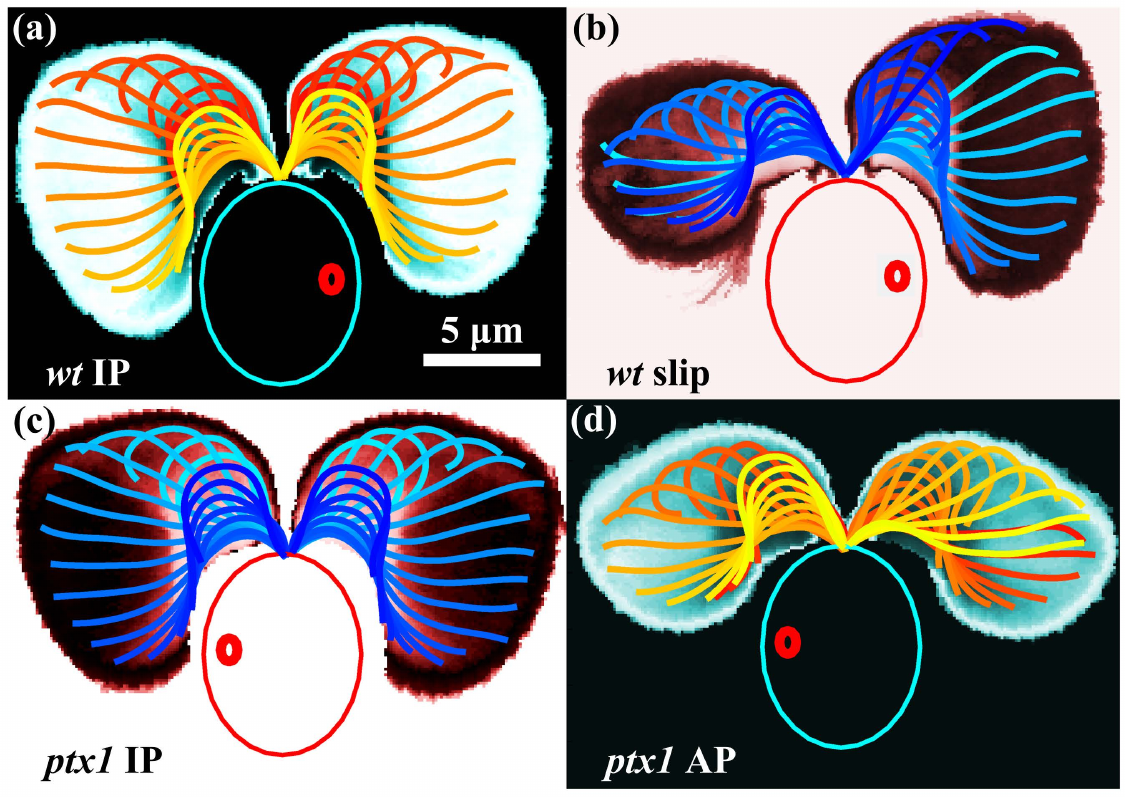}
\caption{(color online).  Waveforms of {\it C. reinhardtii}.  Logarithmically-scaled residence time plots averaged
over ${\cal O}(10^2)$ beats overlaid by waveforms during a cycle, color-coded in time.  The wild-type 
({\it wt}) displays
in-phase (IP) breaststroke beating (a) stochastically interrupted by phase slips (b) in which one flagellum
(here, {\it trans}) beats faster with an attenuated waveform.  The mutant {\it ptx1} displays
an IP state (c) nearly identical to the wild-type (a), and a high-frequency antiphase (AP) state (d).  Large 
and small ovals indicate cell body and eyespot, respectively.}
\label{fig:waveforms}
\end{figure}

Flagellar synchronization is more complex than the 
simplest deterministic models of coupled oscillators would suggest;  beating is
intrinsically stochastic,
cells can switch between synchrony and asynchrony \cite{Polin}, and flagella 
existing within a single organism can be functionally distinct.  
These features are well-established for {\it Chlamydomonas}; the flagella of wild-type ({\it wt}) cells
typically exhibit a noisy in-phase (IP) breaststroke (Fig. \ref{fig:waveforms}a). 
Termed \textit{cis} and 
\textit{trans} for their proximity to the cell's eyespot, the two flagella are 
differentially affected by internal calcium levels, exhibiting a tunable flagellar dominance \cite{dominance}
that allows for phototactic turning. 

\begin{figure*}[t]
\centering
\includegraphics[width=1.80\columnwidth]{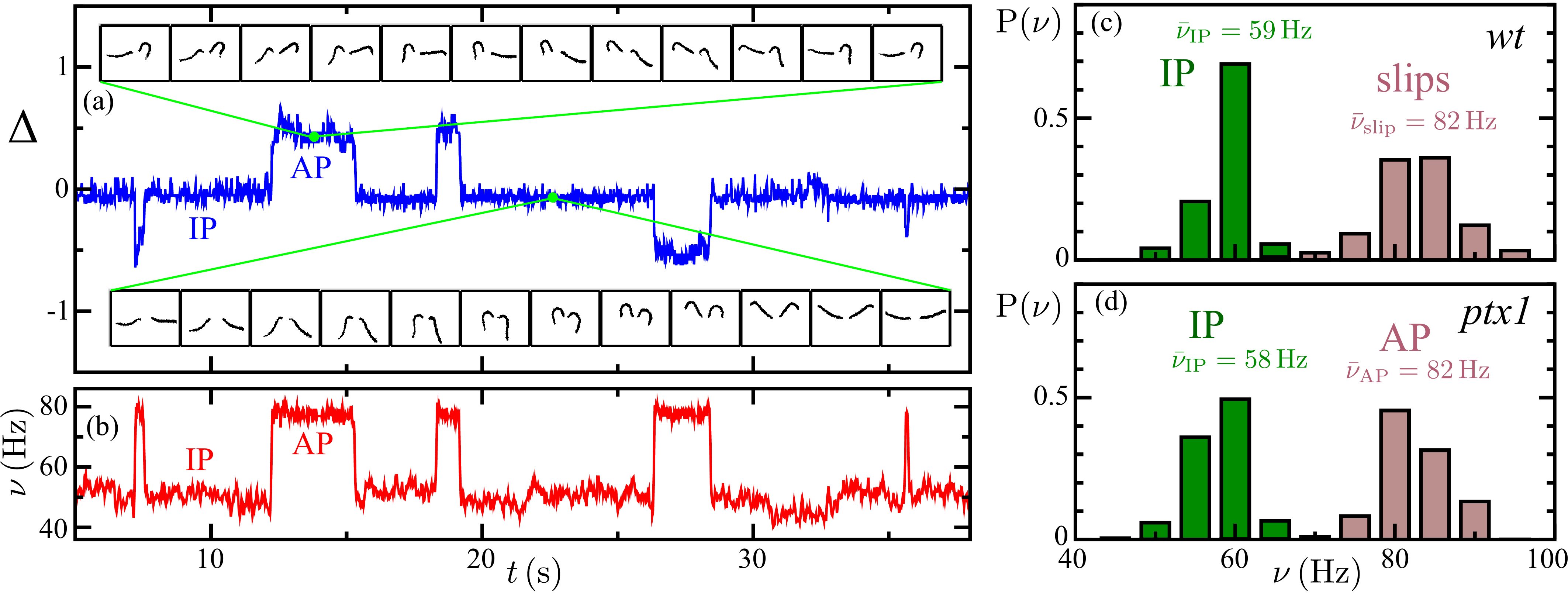}
\caption{(color online).   Beating dynamics.  (a) Phase difference $\Delta=(\psi_{\textit{trans}}-
\psi_{\textit{cis}})/2\pi$ showing half-integer jumps between IP and AP states.  Insets show waveforms in the two states.
(b) Instantaneous frequencies of AP and IP states. (c) Distribution 
of per-beat instantaneous frequencies during IP beating and of the faster flagellum during slips, across all sampled {\it wt} cells. (d) IP and AP per-beat instantaneous frequency distributions, across all sampled {\it ptx1} cells.}
\label{fig:antiphase_seq}
\end{figure*}

We report here an alternative mode of synchronization not previously quantified \cite{RNptx1} in eukaryotes,
in which flagella lock in \textit{antiphase} (AP) synchronization.
For a range of experimental conditions \cite{ptx1long}, this behavior can be sustained in time by 
the ``flagellar-dominance'' mutant \textit{ptx1} of \textit{C. reinhardtii} \cite{Horst1993}. 
While vegetative cells of \textit{ptx1} 
exhibit no gross motility defects, they have defective phototaxis \cite{Horst1993,RNptx1,Okita2005} 
thought to arise from lack of Ca$^{2+}$-dependent flagellar dominance.  
We discuss mechanisms proposed for AP synchronization 
\cite{RaminGeneric,Leoni,Cicuta,Friedrich2012,Ramin3sphere}, and suggest that our 
observations support active filament models
\cite{Camalet} which exhibit discrete
undulating modes of flagella.

Wild-type (CC125) and \textit{ptx1} mutant (CC2894) strains of \textit{Chlamydomonas reinhardtii} \cite{CRC} were grown 
photo-autotrophically in Tris-Minimal medium \cite{Rochaix} with revised trace elements \cite{Kropat} and air 
bubbling in a diurnal growth chamber at 
$24^\circ$C on a 
14:10 h light-dark cycle with a light intensity of 90 $\mathrm{\mu E\,m^{-2}\, s^{-1}}$ \cite{Polin}. Cells were 
harvested from 1 or 2 day-old cultures at a density $\sim \mathrm{6\times10^5}$ cells/ml, during hours 
4 and 5 of the subjective day. Cells were washed in fresh buffer HKC-40/10 (5 mM HEPES, 40 mM KCl, 
10 mM $\mathrm{CaCl_2}$, pH 7.2) and allowed to regrow flagella for at least 2 hours.  Density and 
motility were monitored prior to harvesting and after washing. 	
Specially designed cylindrical PDMS chambers $15$ mm in diameter $\times  4$ mm height were cast 
in custom aluminum molds and plasma-etched onto $22 \times 50$ mm cover slips.  Chambers were placed on a 
Nikon TE2000-U inverted microscope with a $\times63$ Plan-Apo water-immersion objective 
(Carl Zeiss AG, Germany). Micropipettes were used to 
hold and orient cells as described previously \cite{Polin}.
Bright-field illumination was carried out using a halogen lamp with a red long-pass filter ($>$ 620 nm) 
to minimize phototactic 
behavior during experiments, which were performed in the absence of background illumination. 
Video microscopy was performed at $1,000$ fps (Fastcam 
SA3, Photron, USA), post-processed with custom MATLAB code. After each recording the filter was removed 
to locate the orange-colored eyespot and thereby identify the
{\it cis} and {\it trans} flagella. 
Based on the experiments with \textit{wt} cells we concluded 
\textit{Chlamydomonas} need to be acclimated for at least $20-30$ min before characteristic synchronized
breaststrokes are observed \cite{RNall, Polin}. 
Data from $10$ \textit{wt} cells and $12$ \textit{ptx1} cells were analyzed.

There are four key experimental results. The first is the existence of the AP state itself (Fig.~\ref{fig:waveforms}d), 
visualized here by discrete waveforms within one cycle, color-coded in time and overlaid 
on a spatial map of flagellar residence time, averaged over many cycles. 
For reference, compare this to Fig.~\ref{fig:waveforms}a, which shows the {\it wt} IP breastroke waveform. 
Here, the flagella simultaneously execute extended ``power strokes'' followed by 
high-curvature ``recovery strokes'', in which they are drawn forwards with distal portions sliding past the cell body. 
In the AP of the mutant, distinct power and recovery strokes are still clearly discernible, but as one flagellum 
executes the former, the other proceeds through the latter. 
The mutant also displays an IP state (Fig.~\ref{fig:waveforms}c) that is nearly \cite{ptx1long}
identical to the {\it wt} IP.
For example, the areas ${\cal A}^{\it wt,\it ptx1}_{\rm IP}$  swept out by the flagellum in both cases (i.e. the areas within residence-time plots in Fig. \ref{fig:waveforms}) agree to within $1\%$.
In the case of {\it ptx1}, evident also is the drastic reduction in spatial extent spanned 
by both flagella during 
AP relative to the {\it wt} IP mode. 
This alteration of beating {\it waveform} occurs concomitantly with an abrupt increase in 
beating {\it frequency}, 
which together comprise our second observation. 
We extract flagellar phases $\psi_{cis,trans}$ from Poincar{\'e} sectioning of the dynamics as done previously \cite{Polin}, and define the interflagellar phase difference as
$\Delta=(\psi_{trans}-\psi_{cis})/2\pi$.
For a typical {\it ptx1} cell, Fig. \ref{fig:antiphase_seq}a tracks $\Delta(t)$ over $\sim 40$ s. 
We see that $\Delta$ fluctuates around half-integer values during AP, but around integer values during IP. 
As seen in  Fig. \ref{fig:antiphase_seq}, our third finding is precisely that flagella of {\it ptx1} 
stochastically transition back and forth between these IP and AP modes, in a manner reminiscent of the 
synchronous/asynchronous transitions of the {\it wt} \cite{Polin}. 
Fig. \ref{fig:antiphase_seq}b shows that the instantaneous beat frequency is indeed higher in AP ($\nu_{\rm AP}: 82\pm 4$ Hz) than in 
IP ($\nu_{\rm IP}: 58\pm 5$ Hz).   
Fourth and finally, we highlight the striking similarities between the AP state and the state of the flagellum that accumulates one additional cycle during a phase slip of the {\it wt} \cite{Polin}.
This is evidenced by the equivalence both of the waveforms (Fig.~\ref{fig:waveforms}b,d, areas ${\cal A}_{\rm slip}^{\it wt}$,${\cal A}_{\rm AP}^{\it ptx1}$ agree to within $5\%$), and of the frequencies (Fig.~\ref{fig:antiphase_seq}c,d).
The latter figure showing also that {\it wt} and {\it ptx1} cells beat at similar frequencies during IP.

\begin{figure}[t]
\centering
\includegraphics[width=1.0\columnwidth]{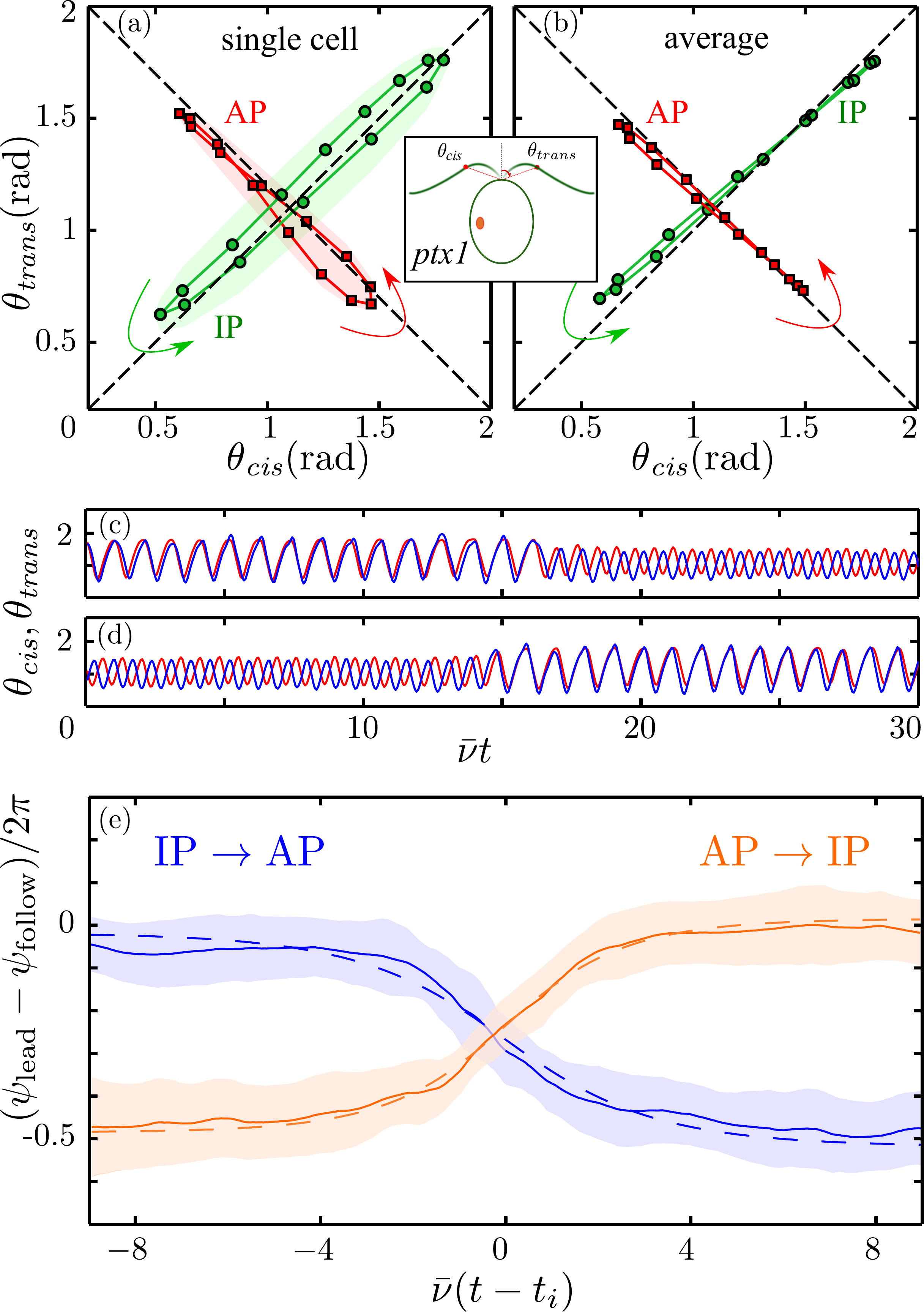}
\caption{(color online).  Synchronization dynamics.  Phase plane of polar angles $\theta_{\it{cis},\it{trans}}$ of a 
single point on each flagellum reveals the IP (green) and AP (red) synchronization 
of a single cell (a), and (b) the average over $6$ cells, averaged over ${\cal O}(10^3)$ beats and
resampled at $15$ points, equally-spaced in time.  Shaded regions in (a) indicate one standard
deviation of stochastic fluctuations. 
(c),(d) are sample timeseries for evolution of $\theta_{\it{cis},\it{trans}}$ during a typical transition event. 
(e) Phase difference dynamics during transitions from
AP to IP (orange) and IP to AP (blue) for $60$ transitions taken from all sampled cells, with means (solid lines) and standard deviations (shaded). Transitions have been vertically aligned by plotting difference modulo $1$. Dashed lines are fits to data.}
\label{fig:phasesep}
\end{figure}

The hypothesis that there is a second, distinct
beating mode of the flagellum is supported by 
estimates of the flagellar force $F$ and power $P$ \cite{three_beats}. 
In a caricature of the power stroke we imagine a straight flagellum of length $L$ pivoting from 
initial polar angle $\theta_0$ to a final one
$\theta_f$ during half the beat period.  
Using resistive force theory we integrate the normal component of the 
viscous force along the filament to obtain $F\sim 2\zeta_{\perp}\nu{\cal A}$, where  
$\zeta_{\perp}$ is the perpendicular drag coefficient per unit length and ${\cal A}$ is the waveform area defined previously.
A similar calculation yields the power $P\sim (2/3)FV$, where $V=L\dot\theta$ is the tip speed of the
flagellum.  Ratios of the product $\nu{\cal A}$ thus serve as measures of relative force in different beats. 
Restricting to a subset of cells whose flagella were most planar, averaged values of the pairs $(\nu,{\cal A})$ for the four states of interest are:
{\it ptx1} IP: ($57.2$ Hz, $147.3$ $\mu$m$^2$),
{\it ptx1} AP: ($81.0$ Hz, $105.1$ $\mu$m$^2$),
{\it wt} IP: ($59.4	$ Hz, $148.8$ $\mu$m$^2$),
{\it wt} slip: ($82.0$ Hz, $110.1$ $\mu$m$^2$).
We find $F^{\it ptx1}_{\rm IP}/F^{\it ptx1}_{\rm AP}=0.99\pm 0.06$ and
$F^{\it wt}_{\rm IP}/F^{\it wt}_{\rm slip}=0.98\pm 0.07$.  
The quantitative match of these ratios supports the identification of a {\it wt} slip with the transient appearance of a higher mode, and the fact that the common value is accurately unity would also imply equal force generation in the two states.
Intriguingly, the ratio of the average AP and IP frequencies 
for {\it ptx1} and of the average slip and IP frequencies of 
the {\it wt} are nearly identical, with a value close to $4/3$. 
Finally, detailed studies \cite{three_beats} show that the peak force during IP power strokes are $\sim 20$ pN
with peak powers ${\cal O}(10 \,{\rm fW})$.  These are in agreement with estimates from 
time-resolved PIV measurements of
energy dissipation in the fluid around free-swimming cells \cite{Guasto}.   

The polar angles ($\theta_{\it cis},\theta_{\it trans}$) measured from the cell midline to equivalent points on 
the two flagella define a 
low-dimensional phase space with which to quantify synchrony.  Figures \ref{fig:phasesep}a,b show IP and AP 
motion in this space for a single cell and a multi-cell average.  Individual cells orbit fairly close to the diagonals, but the mean
displays remarkably precise IP and AP motion, with phase coherence maintained during power and recovery strokes.  
Transitions to and from these two types of synchrony (Figs. \ref{fig:phasesep}c,d) are always initiated by one flagellum, either {\it cis} or {\it trans}, which undergoes alteration of beating mode first \cite{ptx1long}.
Using Poincar{\'e} sections we examine the re-emergence of synchrony during transitions between the two modes using the difference 
$(\psi_{\rm lead}-\psi_{\rm follow})/2\pi$ between the phase of the flagellum that leads the transition and that which follows.
The transition dynamics of respectively AP$\to$IP and IP$\to$AP obey an equivalent functional form derived, on a phenomenological level, from a noisy Adler equation for $\Delta$ \cite{Polin}
\begin{equation}
\dot\Delta=-V'(\Delta)+\xi(t)~.
\label{Adler}
\end{equation}
Here $V(\Delta)=-\delta\nu \Delta + U(\Delta)$, with $\delta\nu$ an intrinsic frequency difference and $U$ 
an effective potential periodic in $\Delta$, and $\xi(t)$ is a noise term.   
Applying this to either type of 
synchrony in {\it ptx1} we
expect $\delta\nu\simeq 0$ due to the lack of flagellar dominance \cite{Okita2005}.  The most parsimonious model would then be
 $U=-\epsilon\cos(2\pi\Delta)$, with
$\epsilon>0$ for AP$\to$IP and 
$\epsilon<0$ for IP$\to$AP.  Solving for the deterministic dynamics ($\xi=0$) in a scaled time 
$s=\nu(t-t_{i})$ centered at the inflection point of the transition $t_{i}$, where $\nu$ is the IP frequency,
we obtain $\Delta=-(1/2\pi)\cos^{-1}\tanh(s/\tau)$, with rescaled relaxation time $\tau=1/(4\pi^2\epsilon/\nu)$.
Fits to the data yield $\tau_{\rm{AP\to IP}}=1.65\pm0.02$ and $\tau_{\rm{IP\to AP}}=-2.07\pm0.04$
(Fig. \ref{fig:phasesep}c,d) and thus
$\epsilon_{\rm{AP\to IP}}/\bar{\nu}\simeq 0.015$ and $\epsilon_{\rm{IP\to AP}}/\bar{\nu}\simeq -0.012$, consistent with the {\it wt} \cite{Polin}.

The necessity to invoke couplings of opposite sign to account for the AP and IP states within the simplest
Adler equation (\ref{Adler}) provides a natural starting point for a discussion of 
mechanisms proposed for synchronization.  Two key issues arise: the structure of the potential $U$ and the origin of the 
coupling constants.    With $\delta\nu=0$, the solution to the Fokker-Planck equation 
for the probability distribution function $P(\Delta)$ associated with (\ref{Adler}) gives $\beta U=-\log[P(\Delta)]$ 
with $\beta$ related to the noise in the usual manner.
The function $\beta U$ so determined \cite{DiLeonardo} will be a bistable potential with 
local minima at integers and half-integers.
This could be accommodated by higher-order Fourier components, 
as $U(\Delta)\simeq -\epsilon\cos(2\pi\Delta)-\alpha\cos(4\pi\Delta)$,
with $\epsilon>0$ and $\alpha>\epsilon/4$.
An alternative to this picture of a {\it fixed} potential landscape $U(\Delta)$ with stochastic hopping between locally-stable 
minima is a 
{\it fluctuating} landscape switching between potentials $U_{IP}$ and $U_{AP}$, the former with minima
only at integers, the latter at half-integers.   Within the limitations of a phase-oscillator description in which 
amplitude dynamics are suppressed, the distinction between these views is fundamentally a 
matter of which degrees of freedom are considered part of the 
dynamical system and the relative time scales for those variables.  In fact,
precedent for a fluctuating landscape can even be seen in the {\it wt} \cite{Polin}, in which 
asynchronous beating (``drifts") 
corresponds to a 
washboard potential tilted by a large $\delta \nu$ so there are no local minima, while synchronous beating has
$\delta\nu$ small enough to allow local minima.  

\begin{figure}[t]
\centering
\includegraphics[width=1.0\columnwidth]{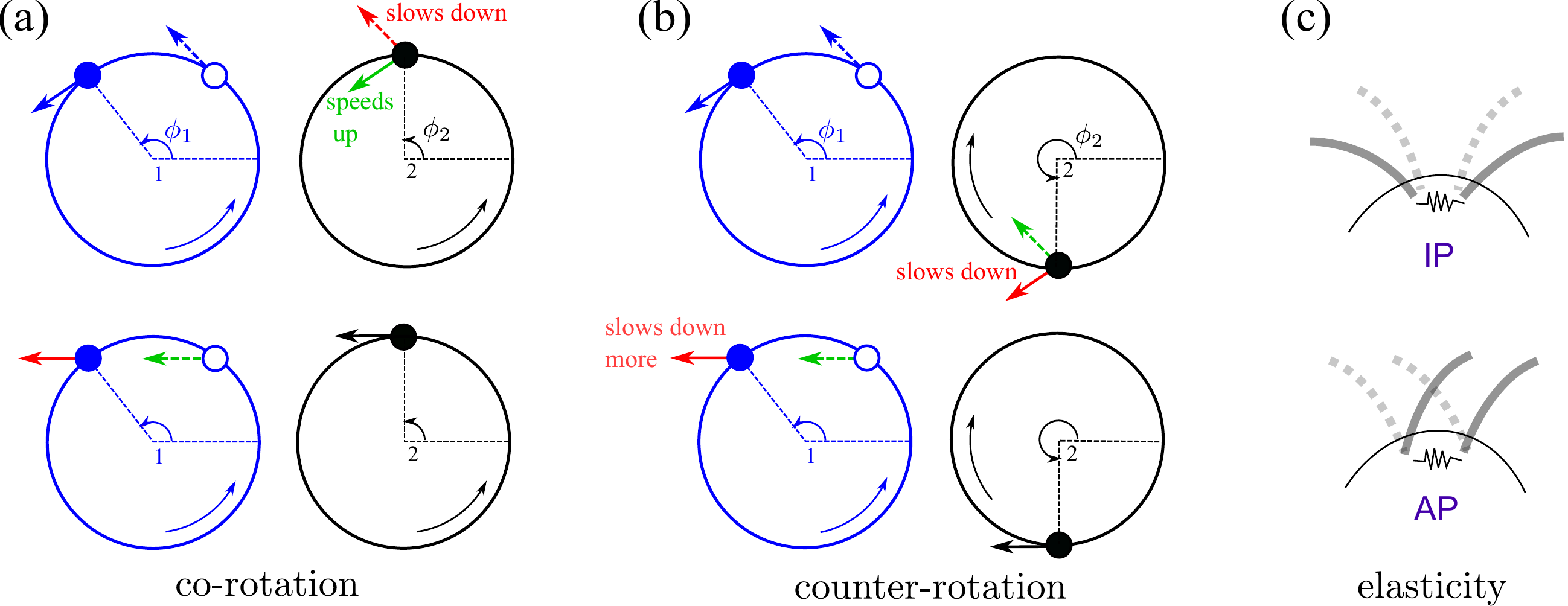}
\caption{(color online). Synchronization mechanisms.  In the elastohydrodynamic model, co-rotating spheres (a) synchronize in phase, while counter-rotating spheres (b) do so in antiphase. 
Top row: motion of sphere $1$ at two possible phases $\phi_1$ induces flows (blue arrows) which 
alter the trajectory of sphere $2$, either speeding it up (green), or slowing it down (red). 
Bottom row: the converse perspective.
(c) Elastic coupling between flagella can induce either IP or AP modes.}
\label{fig:mechanisms}
\end{figure}

Models of synchronization based on hydrodynamic coupling often represent flagella 
by microspheres executing trajectories driven by a tangential internal force.   That force may be considered constant 
along a trajectory with elastic compliance \cite{Niedermayer}, or the trajectories are rigid and the  
forcing varies with orbital phase \cite{RaminGeneric}.  
The mechanism of synchronization in the first class is illustrated in Fig. \ref{fig:mechanisms}a,b.   Measuring
the phase angles ($\phi_1,\phi_2$) of the spheres as indicated, cilia would be modelled by orbits in the
same direction, say $\omega_1\equiv\dot\phi_1>0$ and $\omega_2\equiv \dot\phi_2>0$ (Fig. \ref{fig:mechanisms}a).
If sphere $1$ lags $2$ then the flow produced by $1$ will
push $2$ to a larger radius.  If the internal force is constant, $\dot\phi_2$
will decrease, and $1$ will catch up.  Conversely, if $1$ leads $2$ then it pushes $2$ inward, so $2$ acquires a higher
phase velocity and will catch up. Similarly, the flow induced at $1$
by $2$ leads to consistent results, showing that co-rotating IP motion is stable. 
To model {\it Chlamydomonas} the two spheres must be counter-rotating, with say $\omega_1>0$ and $\omega_2<0$ 
(Fig. \ref{fig:mechanisms}b).  Then, these considerations, together with anisotropy of the stokeslets, predict
stable AP synchronization.  Indeed, the coupling constant in \eqref{Adler} scales as $\epsilon 
\propto -\omega_1\omega_2$ and is negative (positive) for 
co-(counter-)~rotation.   In this simple model the 
AP beating of {\it ptx1} is the `normal' behavior, and the IP mode is anomalous!  The situation is not so clear, though, for  
if the relationship between orbital radius and phase velocity is reversed 
then the coupling changes sign \cite{Leoni,Cicuta}. 
This relationship could be influenced by mechanosensitive cues \cite{Fujiu}.
In the class of models with forcing that varies with phase angle, synchronization can be understood
by similar means in terms of the flow induced by one sphere at the other.
Allowing for non-circular trajectories as well as proximity to a no-slip surface leads to the possibility of an effective potential with the higher-harmonic structure discussed above, stabilizing both IP and AP patterns \cite{RaminGeneric,Ramin3sphere}.
The difficulty in determining the relevance of these arguments to 
{\it ptx1} is precisely that the two modes of synchronization are associated with very distinct waveforms, with potentially different 
compliances, internal forcing, and proximity to the cell surface.   
A third model \cite{Friedrich2012} builds on the fact that transient 
deviations from locked phases will lead to yawing motion of the cell body which 
can produce differential forces on the flagella, bringing them back into phase.
While such a mechanism may pertain to free-swimming cells, it is not immediately clear how it can 
encompass the appearance of both IP and AP states of cells held strongly on micropipettes, where 
we observe only minute angular displacements (below $1^{\circ}$ in both states).
The presence of the cell-body itself appears not to be essential for synchrony of the two flagella, for 
a {\it wt}-like breaststroke gait has been observed in isolated flagellar apparati
(axonemes still connected through their basal bodies), after reactivation by ATP \cite{Hyams}.

No existing models of eukaryotic flagella explain the antiphase 
waveform nor explain why it should emerge.  Approaches based on optimizing swimming efficiency or nutrient 
uptake in a model of {\it Chlamydomonas} \cite{Tam} do find a mode comparable to the IP state.
%A qualitatively similar waveform emerges from a kinematic approach accounting for elastic bending moments \cite{Eloy}, for a single flagellum pumping fluid parallel to a no-slip surface. 
Perhaps the AP waveform is not optimal in any conventional sense, but 
instead exists as one of a discrete number of modes that can emerge from sliding filament models \cite{Camalet}. 
It will be important to establish whether the higher frequency and distinct waveform are properties intrinsic to a single flagellum, or derive from interactions between the two; key insight will likely be gained from examining flagellar dynamics of uniflagellated double mutants of {\it ptx1}.

The physiology of stochastic transitions in the pattern of flagellar beating (i.e. slips or transitions 
in and out of AP) is currently unknown; we hypothesize that fluctuations in the concentration of a 
small molecule or ion might be the origin. One 
candidate would be Ca$^{2+}$, which in isolated and reactivated flagellar axonemes is known to 
control the waveform \cite{cal-waveform}. Interestingly, calcium ions are also responsible for 
the contractility of striated fibers that connect the 
basal bodies of the two flagella \cite{cal-fibers}, which in turn may act as a spring with 
variable stiffness. The current state of this potential spring may influence the preferred 
mode of synchronization. Indeed, generalizing the 
orbiting-sphere model \cite{Niedermayer} to include
an elastic connection between flagella bases can lead to stabilization of either the
IP or AP modes (Fig. \ref{fig:mechanisms}c), depending on microscopic details of the elasticity. 
In the simplest linear
spring, for example, the AP mode (termed `parallel' by R{\"u}ffer and Nultsch \cite{RNptx1}) can be selected, 
for it is the mode in which the relative displacements of the flagellar connections within the cell body are most nearly constant.
The role of these fibers for flagellar synchronization may be clarified by altering their mechanical properties by chemical or other means.

KCL and KYW contributed equally to this work. We thank D. Page-Croft for technical assistance. Support is acknowledged
 from the Spanish Ministerio de Ciencia y Innovaci\'on grant FIS2010-22322-C01 and a Ram\'on y Cajal Fellowship (IT), an 
EPSRC postdoctoral Fellowship (MP), the BBSRC, the EPSRC, ERC Advanced Investigator 
Grant 247333, and a Senior Investigator Award from the Wellcome Trust (REG).

%%%%%%%%%%%%%%%%%%%


\begin{thebibliography}{99}

\bibitem{Collins1993} J.J. Collins and I.N. Stewart, J. Nonlinear Sci. {\bf 3}, 349 (1993).

\bibitem{LaugaGoldstein2012} E. Lauga and R.E. Goldstein, Physics Today {\bf 65}, 30 (2012).

\bibitem{chlamy_sourcebook} E. H. Harris, {\it The Chlamydomonas Sourcebook} (AcademicPress, Oxford, 2009), Vols 1,3.

\bibitem{RNall} U. Ruffer, W. Nultsch, Cell Motil. Cytoskeleton \textbf{5}, 251 (1985); \textbf{7}, 87 (1987). 

\bibitem{Polin} M. Polin {\it et al.}, Science \textbf{325}. 487 (2009); R.E. Goldstein, M. Polin, I. Tuval, 
Phys. Rev. Lett \textbf{103} 168103 (2009); {\bf 107}, 148103  (2011). 

\bibitem{KnightJones} E.W. Knight-Jones, Q. J. Microsc. Sci. {\bf 95}, 503 (1954).

\bibitem{Brumley} D.R. Brumley, M. Polin, T.J. Pedley, and R.E. Goldstein, Phys. Rev. Lett. {\bf 109}, 268102 (2012).

\bibitem{RaminGeneric} R. Golestanian, J.M. Yeomans, N. Uchida, Soft Matter \textbf{7}, 3074 (2010); 
N. Uchida, R. Golestanian, Phys. Rev. Lett. \textbf{106} 058104 (2011); N. Uchida, R. Golestanian, Eur. Phys. J. E (2012)

\bibitem{Niedermayer} T. Niedermayer, B. Eckhardt and P. Lenz, Chaos {\bf 18}, 037128 (2008).

\bibitem{Guirao} B. Guirao and J.-F. Joanny, Biophys. J. 92, 1900
(2007).


\bibitem{dominance} K. Yoshimura, Y. Matsuo, R. Kamiya, Plant Cell Physiol. \textbf{44}. 1112 (2003) 

\bibitem{RNptx1} U. Ruffer, W. Nultsch, Cell Motil. Cytoskeleton. \textbf{37}, 111 (1997);  
\textbf{41}, 297 (1998).  In these works, what we term the AP and IP states were described 
qualitatively from light-table tracings of frames from short high-speed movies. 

\bibitem{ptx1long} K.C. Leptos, K.Y. Wan, and R.E. Goldstein, preprint (2013).

\bibitem{Horst1993} C. Horst, G. Witman, J. Cell. Biol. \textbf{120}, 733 (1993). 

\bibitem{Okita2005} N. Okita, N. Isogai, M. Hirono, R. Kamiya, K. Yoshimura, J. Cell. Sci. \textbf{118} 529 (2005).

\bibitem{Leoni} M. Leoni and T.B. Liverpool, Phys. Rev. E {\bf 85}, 040901 (2012).

\bibitem{Cicuta} N. Bruot, J. Kotar, F. de Lillo, M. Cosentino Lagomarsino, and P. Cicuta, Phys. Rev. Lett. {\bf 109}, 164103 (2012).

\bibitem{Friedrich2012} B.M. Friedrich and F. J\"ulicher, Phys. Rev. Lett. \textbf{109} 138102 (2012).

\bibitem{Ramin3sphere} R.R. Bennett, R. Golestanian, Phys. Rev. Lett. {\bf 110}, 148102 (2013).

\bibitem{Camalet} S. Camalet and F. J{\"u}licher, New J. Phys. {\bf 2}, 24 (2000).

\bibitem{CRC} Chlamydomonas Resource Center at the University of Minnesota, http://www.chlamy.org.

\bibitem{Rochaix} Rochaix J.-D., Mayfield S., Goldschmidt-Clermont M. and Erickson J.M.: Molecular biology of 
Chlamydomonas. In: Plant molecular biology: a practical approach. 1988, pp. 253-275. Ed. by Schaw C.H. IRL Press (Oxford). 

\bibitem{Kropat} J. Kropat {\it et al.} Plant J. {\bf 6} 770 (2011).

\bibitem{three_beats}   Fully time-resolved measurements are discussed  
in: K.Y. Wan, K.C. Leptos, and R.E. Goldstein, preprint (2013).

\bibitem{Guasto} J. S. Guasto, K. A. Johnson, and J. P. Gollub, Phys. Rev. Lett. {\bf 105}, 168102 (2010).

%\bibitem{Kuramoto_delays} H.G. Schuster and P. Wagner, Prog. Theor. Phys. {\bf 81}, 939 (1989).

\bibitem{DiLeonardo} R. DiLeonardo {\it et al.}, Phys. Rev. Lett. {\bf 109}, 034104 (2012).

%\bibitem{Drescher2010} K. Drescher, R.E. Goldstein, N. Michel, M. Polin, I. Tuval, Phys. Rev. Lett. \textbf{105} 168101 (2010).

%\bibitem{structbio} C.F. Barber, T. Heuser, B.I. Carbajal-Gonzalez, V.V. Botchkarev, D. Nicastro, Mol. Bio. Cell. \textbf{23} 111 (2012); D. Nicastro, X.F. Fu, T. Heuser, A. Tso, M.E. Porter, R.W. Linck, Proc. Natl. Acad. Sci. U.S.A. \textbf{108} E845 (2012).

%\bibitem{Hale2006} M.E. Hale, R.D. Day, D.H. Thorsen, M.W.Westnear, J. Exp. Biol. \textbf{209}, 3708 (2006).

\bibitem{Fujiu} K. Fujiu, et al., Nat. Cell Biol. {\bf 13}, 630 (2011).

\bibitem{Hyams} J.S. Hyams and G.G. Borisy, J. Cell Sci. {\bf 33}, 235 (1978).

\bibitem{Tam} D. Tam and A.E. Hosoi, Proc. Natl. Acad. Sci. U.S.A. {\bf 108}, 1001 (2011).

\bibitem{cal-waveform} M. Bessen, R.B. Fay, and G.B. Witman, J. Cell. Biol. \textbf{86}. 446 (1980).

\bibitem{cal-fibers} K.-F. Lechtreck, and M. Melkonian, Protoplasma \textbf{164}. 38 (1991).


\end{thebibliography}
\end{document}